\documentclass[aps,twocolumn,floatfix, showkeys, superscriptaddress, nofootinbib]{revtex4}
\usepackage{amsmath}
\usepackage{graphicx,bm,hhline}
\usepackage{subfigure}

\newcommand{\br}{{\bm r}}
\newcommand{\brp}{{\br}^\prime}
\newcommand{\hf}{\frac{1}{2}}

\newcommand\sss{\scriptscriptstyle}

\newcommand{\wig}[1]{\mathrel{\hbox{\hbox to 0pt{\lower.6ex\hbox{$\sim$}\hss}\raise.4ex\hbox{$#1$}}}}

\begin{document}

\title{Coulomb Log for Conductivity of Dense Plasmas}

\author{C. E. Starrett}
\email{starrett@lanl.gov}
\affiliation{Los Alamos National Laboratory, P.O. Box 1663, Los Alamos, NM 87545, U.S.A.}

\date{\today}
\begin{abstract}
The Coulomb log ($\log\Lambda$) approximation is widely used to approximate electron transport
coefficients in dense plasmas.  It is a classical approximation to the momentum
transport cross section.  The accuracy of this approximation for electrical
conductivity in dense
plasmas is assessed by comparing to fully quantum mechanical calculations for realistic
scattering potentials.  It is found that the classical approximation is accurate
to $\pm$10\% when $\log\Lambda > 3$, irrespective of the plasma species.  The
thermodynamic regime (density and temperature) for which $\log\Lambda > 3$ corresponds
to does, however, strongly depend on the material.  For increasing Z, $\log\Lambda$ is
greater than 3
for increasingly high temperatures and lower densities. 
\end{abstract}
\pacs{ }
\keywords{Electrical conductivity, dense plasmas, Coulomb Log}
\maketitle

\section{Introduction}
The Coulomb Log is a widely used concept in low density plasma physics that
allows rapid calculation of the electron-ion momentum transport 
cross section \cite{cohen50}.  It is used to calculate electron and ion transport
coefficients.  The core approximation is that Rutherford's classical
cross section is assumed, but with physically motivated impact parameter
cut-offs to guarantee finite results.  It is also used for modeling dense plasmas
that are found in neutron and white dwarf stars \cite{daligault09, potekhin99}, 
for modeling Inertial Confinement Fusion \cite{lindl95,hu14}, and for
studying temperature equilibration \cite{gericke02, kodanova18, benedict17, daligault09a}
and stopping power \cite{zwicknagel99, morawetz96}.

The dense plasma environment raises questions about the validity of the core approximation:
the classical Rutherford cross section.  This neglects electron screening (polarization),
core-valence orthogonality, and ion correlations; effects that are known to be important
when modeling dense plasmas.  Authors typically use the impact parameter
cut-offs to account for plasma effects such as degeneracy, ion-correlations and electron
screening \cite{lee84, stygar02, filippov18}.  However, such {\it ad hoc} corrections
must be tested.
The accuracy of the Coulomb Log relative to quantum calculations that
assume a Debye screening potential has been tested \cite{gericke02}, as has
the effect of ion-ion correlations
again assuming a Debye interaction \cite{filippov18}. 

In this work the accuracy of the Coulomb Log approach in dense plasmas is assessed through 
comparison with fully quantum mechanical calculations for a scattering
potential that realisitically accounts for all the above mentioned physics.  This 
potential is the so-called potential of mean force \cite{starrett17}.
It is based on density functional theory (DFT) \cite{mermin65} and the quantum 
Ornstien Zernike equations \cite{chihara91}.  Recently it was shown that calculations of
the electrical conductivity using this potential agreed well with existing experiments
and with multi-center DFT MD (molecular dynamics) calculations based on the Kubo-Greenwood
formula \cite{starrett17}.  Hence, through a comparison of electrical conductivity
resulting from the mean force and Coulomb Log approaches the accuracy of the latter is
assessed.

In this work, the accuracy of the Coulomb log
approach is assessed as a function of plasma conditions and make-up (temperature, density and
nuclear charge).  Plasmas of pure hydrogen, beryllium, aluminum, copper, silver and
lutetium are considered, and plasma conditions from 10 eV through 1 keV are covered.
The relationship between the value of the Coulomb log and its accuracy
is also tested for these plasmas.  Since the Classical Coulomb log method relies on an 
input of the average ionization,
a simple and widely used model for estimating the average ionization 
(the Thomas-Fermi Cell model \cite{feynman}) is also tested.

\section{Theory}
\subsection{The Coulomb Log approximation \label{sec_cl}}
The non-relativistic quantum mechanical expression for the momentum transport cross section is
\begin{equation}
\sigma_{\sss TR}(\epsilon) = \frac{4 \pi}{p^2} \sum\limits_{l=0}^{\infty} (l+1) \left( \sin\left( \eta_{l+1} - \eta_{l} \right)\right)^2
\label{trq}
\end{equation}
where the $\eta_l = \eta_l(\epsilon)$ are the scattering phase shifts, $\epsilon$ is the electron 
energy and $p$ its momentum.  The sum over orbital angular momentum quantum
number $l$, while formally to $\infty$ does converge for finite $\epsilon$,
since the phase shifts approach zero for large $l$.
For classical electrons the momentum transport cross section is \cite{morawetz96}
\begin{equation}
\sigma_{\sss TR}(\epsilon) = \frac{4 \pi \bar{Z}^2 e^2}{p^4} \log \Lambda
\label{trc}
\end{equation}
where the average ionization of the plasma is $\bar{Z} = \bar{n}_e^0 / n_i^0$ 
($\bar{n}_e^0$ is the average ionized electron number density, $n_i^0$
the number density of nuclei).
$\log \Lambda$ is the so-called coulomb log \cite{lee84, spitzer40, temko57}
\begin{equation}
\log \Lambda = \hf \log(1+b_{max}^2 / b_{min}^2)
\label{ll}
\end{equation}
This arises from consideration of Coulomb collisions, assuming Rutherford cross 
section and small angle scattering, with $b_{min}$ and $b_{max}$ being the assumed
minimum and maximum impact parameters.  

$b_{max}$ is be assumed to be \cite{lee84}
\begin{equation}
b_{max} = \max(\lambda_{\sss DH}, R)
\label{bmax}
\end{equation}
where $R$ is the ion sphere radius and
$\lambda_{\sss DH}$ is the degeneracy corrected Debye-H\"uckel screening length
\begin{equation}
\lambda_{\sss DH}^{-2} = 
\frac{4\pi \bar{n}_e^0 e^2 }{k(T^2 + T_F^2)^\hf}
+ 
\frac{4\pi \bar{Z}^2 n_i^0 e^2}{kT}
\label{ldh}
\end{equation}
where $T$ is the temperature and $T_F$ is the Fermi temperature.
$b_{min}$ is assumed to be \cite{lee84}
\begin{equation}
b_{min} = \max(\lambda_{\sss dB}, b_\bot)
\label{bmin}
\end{equation}
where $\lambda_{\sss dB}$ is the radian de Broglie wave length 
\begin{equation}
\lambda_{dB} = \frac{\hbar}{2m\bar{v}}
\end{equation}
and $b_\bot$ is the classical closest distance of approach
\begin{equation}
b_\bot = \frac{\bar{Z}e^2}{m\bar{v}^2}
\end{equation}
Here $\bar{v}$ is taken to be 
\begin{equation}
\bar{v} = \max(v_{rms}, v_F)
\label{vbar}
\end{equation}
where $v_{rms}=\sqrt{3kT/m}$, where $T$ is the temperature, is
the root mean squared velocity and $v_F$ is the Fermi velocity.

It is also possible to evaluate the momentum transport cross section in Born
approximation, which is quantum mechanical, but is only correct in the
high electron-energy limit
\begin{equation}
\sigma_{\sss TR}^{\sss Born}(\epsilon) = \frac{\pi}{p^2} \int\limits_{0}^{2p} dq\,q^3 \left| \frac{V(q)}{2\pi} \right|
\label{trb}
\end{equation}
where $V(q)$ is the scattering potential.  If $V(r)$ is assumed to be a Debye potential
then the Born result becomes
\begin{equation}
\sigma_{\sss TR}^{\sss B}(\epsilon) = \frac{4 \pi \bar{Z}^2 e^2}{p^4} \log \Lambda^{\sss B}
\label{trbd}
\end{equation}
where
\begin{equation}
\begin{split}
\log \Lambda^B =& \hf 
\left( \log\left(1+b_{max}^2 / b_{min}^2 \phantom{\hf}\right)   \right. \\
&\left.- \frac{\left(b_{max} / b_{min}\right)^2}{1+ \left(b_{max} / b_{min}\right)^2 } \right)
\label{llb}
\end{split}
\end{equation}
where $b_{max} = \lambda_{\sss DH}$ and $b_{min} = \lambda_{\sss dB}$.  For the Born approximation
the form of $b_{max}$ and $b_{min}$ are not assumed, as they are in the classical case, but
derived.  However, one can view equation (\ref{llb}) as a quantum correction to the classical
formula (\ref{ll}), and use equations (\ref{bmin}) and (\ref{bmax}) for $b_{min}$ and $b_{max}$.

In the results section calculations using equations (\ref{trq}), (\ref{trc}) and (\ref{trbd})
are compared using equations (\ref{bmin}) and (\ref{bmax}) in the classical and Born
approximations.

\subsection{Conductivity Model\label{sec_cm}}
In the relaxation time approximation, or Krook model \cite{bhatnager54, krall73}, the dc conductivity is given by
\begin{equation}
\sigma_{DC} = \frac{1}{3\pi^2} \int_0^\infty \left( -\frac{df(\epsilon, \mu)}{d\epsilon} \right)  v^3 \tau_\epsilon d\epsilon
\label{sdc}
\end{equation}
where $f(\epsilon,\mu)$ is the Fermi-Dirac occupation factor, $\epsilon$ the electron energy $\epsilon = m v^2 /2$, 
$\mu$ is the electron chemical potential and $\tau_\epsilon$ is the relaxation time.
This last is related to the electron-ion momentum transport cross section $\sigma_{TR}(\epsilon)$
\begin{equation}
\tau_\epsilon = \frac{1}{n_i^0\,v\, \sigma_{TR}(\epsilon) }
\end{equation}
The model, equation (\ref{sdc}), ignores the effect of electron-electron collisions.  These can be taken into account,
for example using the method of reference \cite{reinholz15}.  Since our aim here is to assess the accuracy of the 
Coulomb logarithm approach that approximates the electron-ion cross section, it is safe to ignore electron-electron
collisions, with the understanding that if the actual conductivity is required, for example, to compare with experiment,
then this effect must be accounted for.

\begin{figure}
\begin{center}
\includegraphics[scale=0.40]{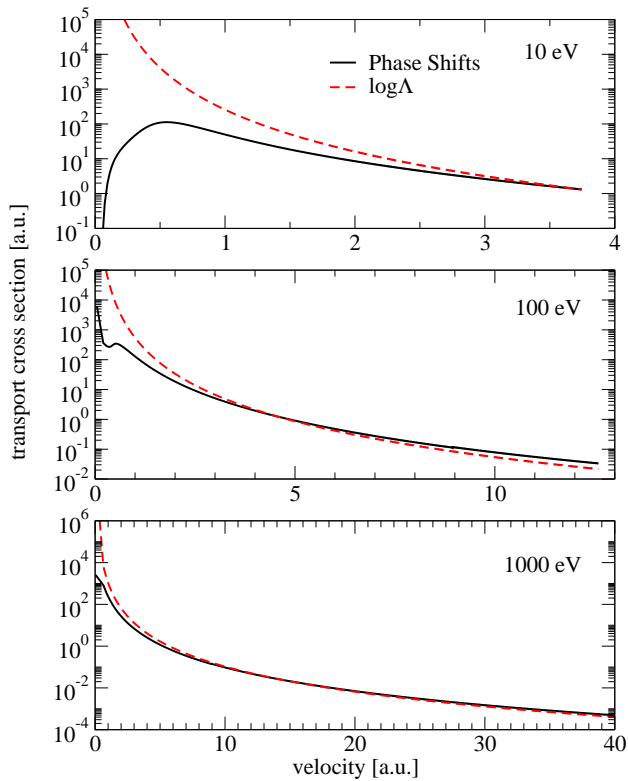}
\end{center}
\caption{(Color online) 
Momentum transport cross sections for beryllium at 0.185 g/cm$^3$.
}
\label{fig_be_tr}
\end{figure}
\begin{figure}
\begin{center}
\includegraphics[scale=0.40]{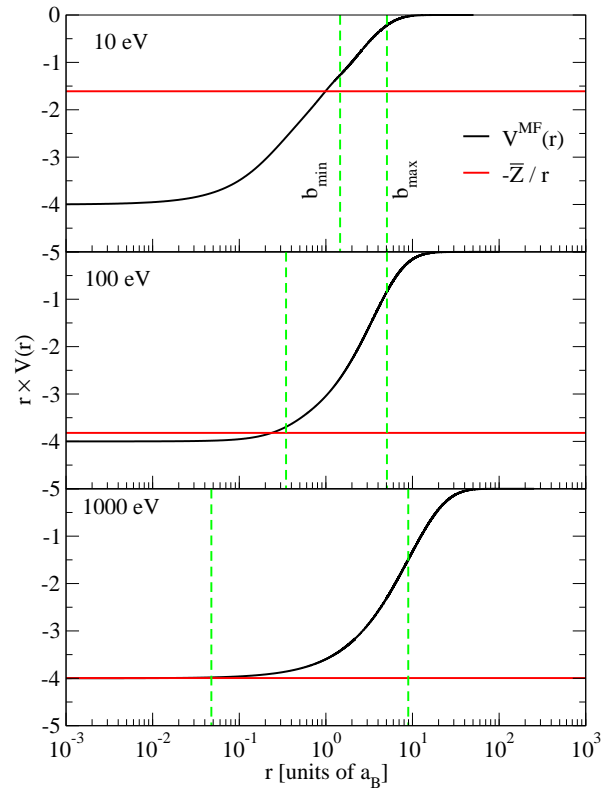}
\end{center}
\caption{(Color online) 
Scattering potentials for beryllium at 0.185 g/cm$^3$.
}
\label{fig_be_v}
\end{figure}

\subsection{Potential of Mean Force Model \label{sec_pm}}
The potential of mean force $V^{\sss MF}(r)$ for conductivity calculations was introduced in
reference \cite{starrett17} in analogy with the well known classical potential of mean force.
It is given by
\begin{eqnarray}
V^{\sss MF}(r) & = & V_{ie}(r) + n_i^0 \int d^3 r^\prime \frac{ C_{ie}(|\br -\brp|) }{-\beta} h_{ii}(r^\prime) \nonumber\\
               &   & + \bar{n}_e^0 \int d^3 r^\prime \frac{ C_{ee}(|\br -\brp|) }{-\beta} h_{ie}(r^\prime) \nonumber\\
\label{vmf1}               
\end{eqnarray}
where $V_{ie}(r)$ is the electron-ion interaction potential, including a nuclear Coulomb term and the Coulombic and exchange electron
interaction terms, $C_{ie}(r)$ ($C_{ee}(r)$)
is the electron-ion (electron-electron) direct correlation function, and $\beta = 1/ kT$, where $T$ is the temperature.

\begin{figure}
\begin{center}
\includegraphics[scale=0.35]{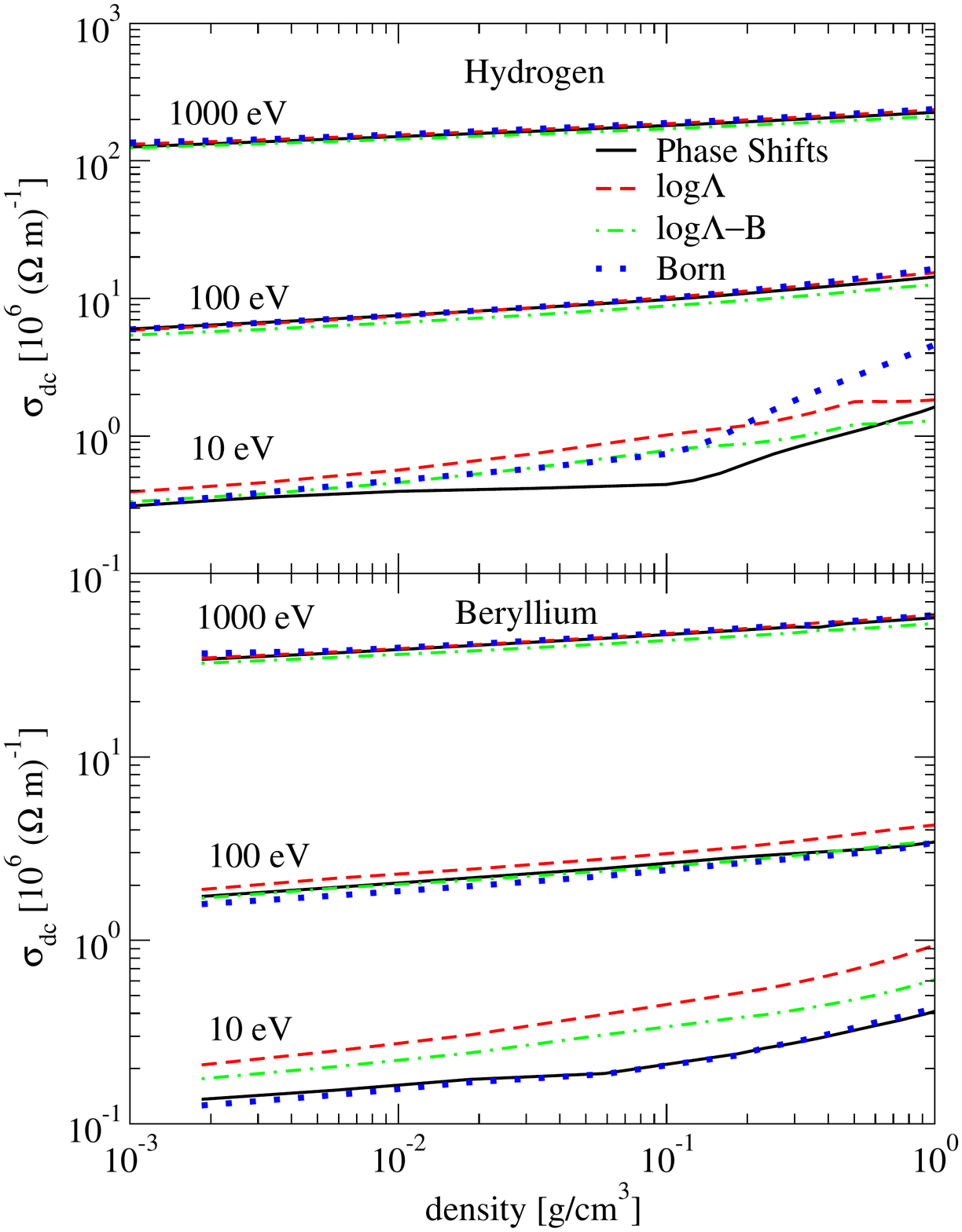}
\end{center}
\caption{(Color online) 
Electron-ion electrical conductivity of hydrogen and beryllium.  
}
\label{fig_hbe}
\end{figure}
\begin{figure}
\begin{center}
\includegraphics[scale=0.35]{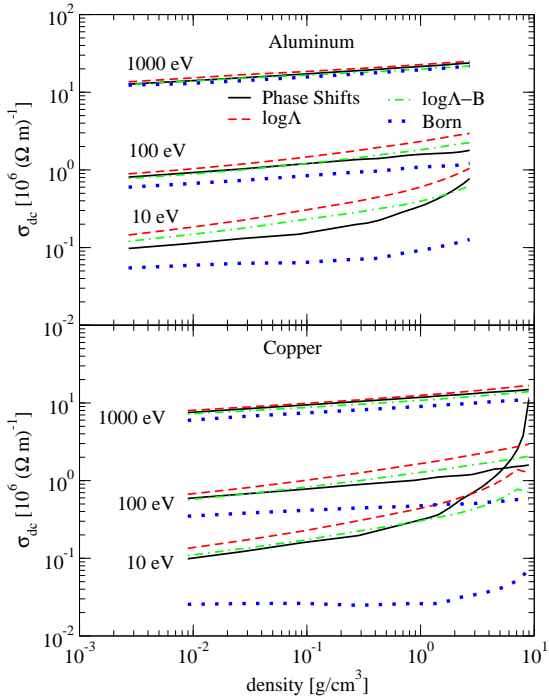}
\end{center}
\caption{(Color online) 
Electron-ion electrical conductivity of aluminum and copper.  
}
\label{fig_alcu}
\end{figure}
\begin{figure}
\begin{center}
\includegraphics[scale=0.35]{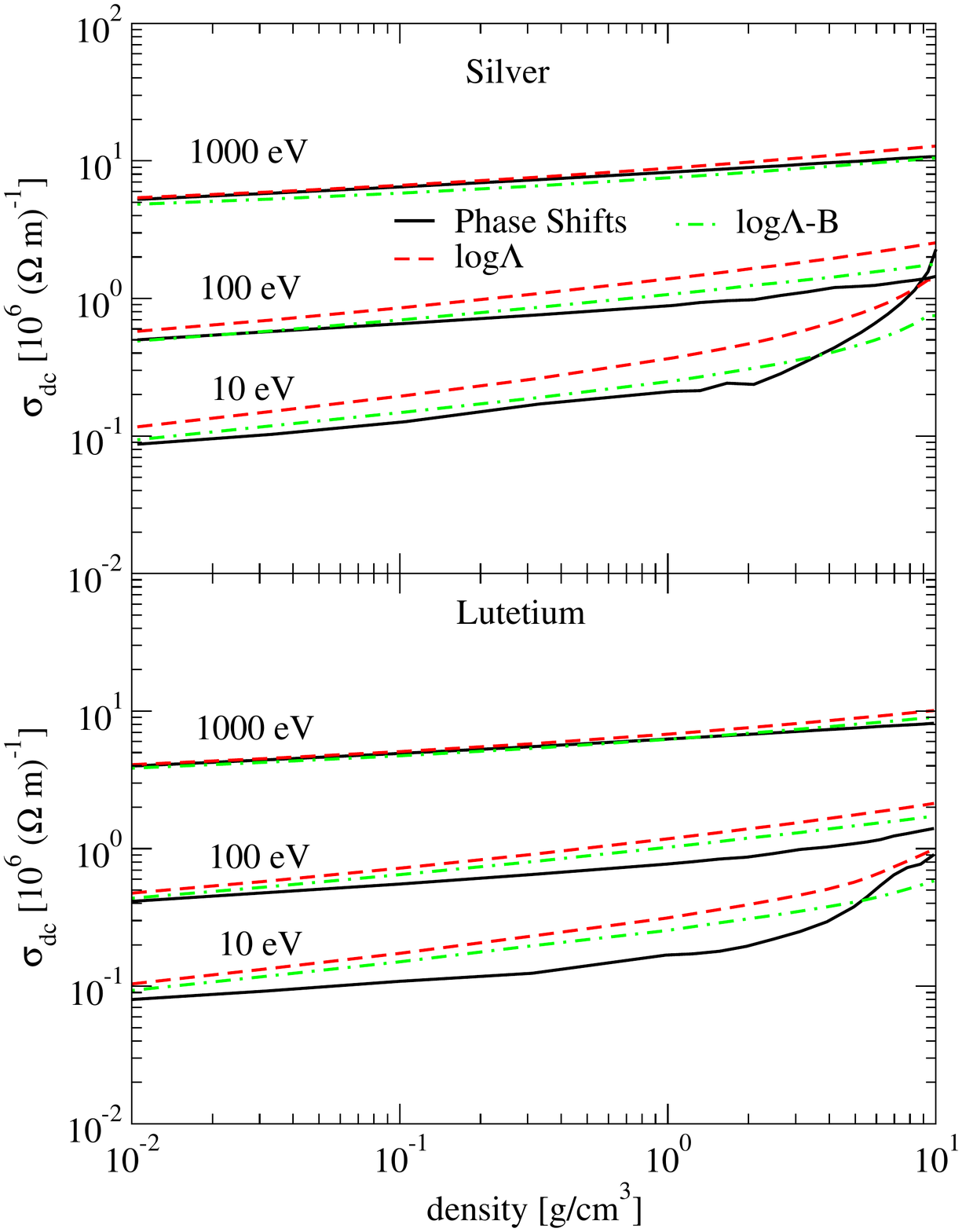}
\end{center}
\caption{(Color online) 
Electron-ion electrical conductivity of silver and lutetium.  
}
\label{fig_aglu}
\end{figure}

The accuracy of conductivity calculations based on this potential was assessed in reference 
\cite{starrett17}.  It was found to be in generally good agreement with available 
experiments, as well as with DFT-MD and Quantum Lenard Balescu calculations \cite{desjarlais17}.  It
was shown to be less reliable in the metal to insulator transition region at low
temperature.  In this work, the focus is on regimes far from this region, i.e.
at higher temperatures.  

The potential is calculated using a DFT based average atom model coupled to the
quantum ornstein-Zernike equations \cite{starrett13, starrett14}.  This model
is well tested and accurate in the dense plasma regime \cite{souza14, starrett15b, daligault16}.
Note that this model also provides the average ionization $\bar{Z} = \bar{n}_e^0 / n_i^0$,
from which $\mu$ is known.  The temperature dependent exchange and correlation
potential of reference \cite{groth17} has been used in all our calculations.

An important modelling assumption is that the conductivity can be calculated
assuming that the plasma is made up of an ensemble of identical pseudo-atoms,
each with an ionization equal to the average ionization of the plasma ($\bar{Z}$).
This latter is determined within the model \cite{starrett14}.  In reality
the plasma is composed of a distribution of ions with different electronic
structures.  This single-average-ionization assumption is not tested here, but
it is noted that this assumption is also used in the classical coulomb log approach, so a comparison
is meaningful.

\section{Results}
The classical model for $\sigma_{\sss TR}$ described in section \ref{sec_cl} is identical to
widely used Lee-More model (LM) \cite{lee84}.  LM also introduce a minimum allowed value for
$\log\Lambda$ of 2.0 \footnote{LM also used a separate model for for solids, a region of phase
space not encountered in this work.}.  Desjarlais \cite{desjarlais01} built on and improved upon
the LM model to make it more accurate for metal-insulator transitions, a regime not
considered here.

In figure \ref{fig_be_tr} the momentum transport cross section for beryllium at 1/10$^{th}$ of
solid density for 10, 100 and 1000 eV is shown.  The fully quantum mechanical calculation
(labeled ``Phase Shifts'') is compared to the classical result (labeled ``$\log\Lambda$'').
The range of velocities shown for each temperature is restricted to the region where
$\sigma_{\sss TR}$ contributed appreciably to the integral in equation (\ref{sdc}).

At 1000 eV the quantum and classical transport cross sections are close due to the high energies
of the electrons.  For small electron velocities the results diverge as the quantum result
is finite, while the classical cross section scales as $\sim v^{-3}$.
For $kT$ = 100 eV, the results are again similar.  But for $kT$ = 10 eV,
there is a very significant difference.  Clearly this is due to the
enhanced importance of low energy electrons relative to the higher
temperature cases.

For the same plasmas as in figure \ref{fig_be_tr}, in figure \ref{fig_be_v}
the scattering potentials are shown.  In the classical case, 
the Rutherford Cross section is for a purely Coulombic potential $-\bar{Z}/r$.
For the quantum case, where $V^{\sss MF}(r)$ is used, the potential
is screened (i.e. short ranged) and $V^{\sss MF}(r) \to -Z/r$ as $r \to 0$, where $Z$ is
the nuclear charge.
Also shown in the figure are $b_{min}$ and $b_{max}$ for each case.  Clearly
the proximity of the transport cross sections in figure \ref{fig_be_tr} is
not due to the similarity of the scattering potentials, but rather the
nearly free electron character of the high energy electrons.

In figures \ref{fig_hbe} to \ref{fig_aglu} the electrical
conductivity of hydrogen, beryllium, aluminum, copper, silver
and lutetium plasmas is shown. The range of densities considered
is $1/1000^{th}$ of normal (i.e solid) to normal density.  For
hydrogen 0.001 to 1 g/cm$^3$ is used.
The results shown in these plots include fully quantum (``Phase Shifts''),
classical (``$\log\Lambda$''), classical but with the Born
correction (``$\log\Lambda-B$'') and quantum Born (``Born'') that
uses equation (\ref{trb}).

The trends are clear;  for ``Born'', agreement with the quantum calculation
is reasonable for low Z (H and Be), but becomes poor for mid Z (Al and Cu).
For high Z (Ag and Lu) the Born result is not shown as it is in so poor agreement
that the figure becomes hard to read.  The Born approximation assumes
that the scattered electron wave function is a plane wave, i.e. a free
state.  Clearly the stronger the scatterer, the worse the Born approximation
will be, as borne out by the results.  Generally speaking, the Born
cross section overestimates the cross section.  Here that translates
to a larger transport cross section, a shorter relaxation time and therefore
a reduced conductivity.

For ``$\log\Lambda$'' and ``$\log\Lambda-B$'' generally excellent agreement
with the quantum calculation is seen
at $kT$ = 1000 eV.  The Born correction sometime makes agreement better,
and sometimes worse, though the effect is small.  At $kT$ = 100 eV, for low Z agreement with the quantum
calculation is good, but becomes poorer as Z is increased.  The Born
correction mostly improves the results, but not universally.  At $kT$ = 10 eV, agreement
of both the $\log\Lambda$ approachs with the fully quantum results is generally relatively poor.

That the Born correction typically (but not universally) improves agreement
with the quantum results is prima facie surprising, given how poor the
quantum Born approximation does, in contrast to the classical ``$\log\Lambda$''.  
However, the classical cross section
without the ad hoc impact parameter cut off's would yeild an infinite cross
section and therefore zero conductivity everywhere.  Hence it is reasonable
that the Born correction could improve agreement.

\begin{figure}
\begin{center}
\includegraphics[scale=0.3]{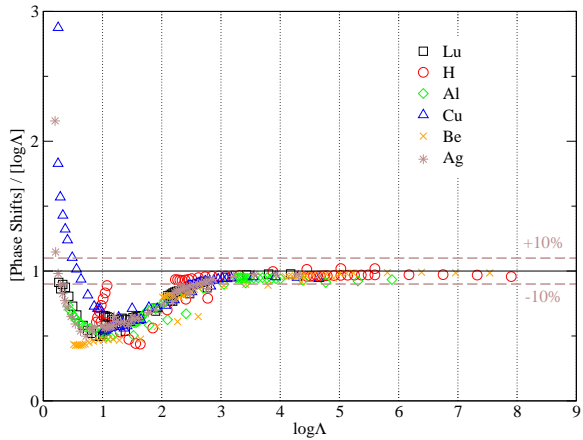}
\end{center}
\caption{(Color online) 
The ratio of the full quantum calculation to the classical $\log\Lambda$ 
approach. 
}
\label{fig_logl}
\end{figure}

\begin{figure}
\begin{center}
\includegraphics[scale=0.3]{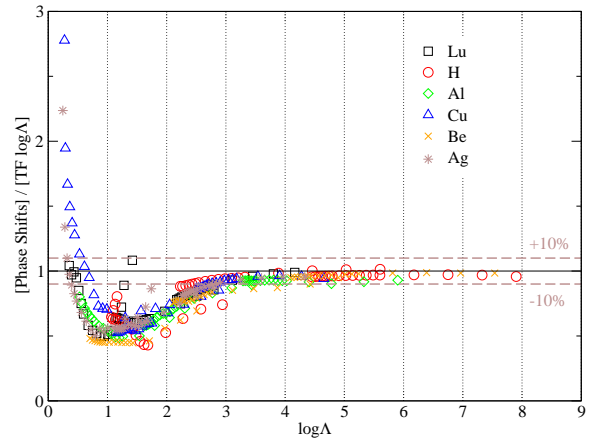}
\end{center}
\caption{(Color online) 
As in figure \ref{fig_logl} but with the ionization for the classical
$\log\Lambda$ method determined using the Thomas-Fermi cell model \cite{feynman}.
}
\label{fig_logl_tf}
\end{figure}
\begin{figure}
  \centering
  \begin{subfigure}[]{} 
    \includegraphics[scale=0.6]{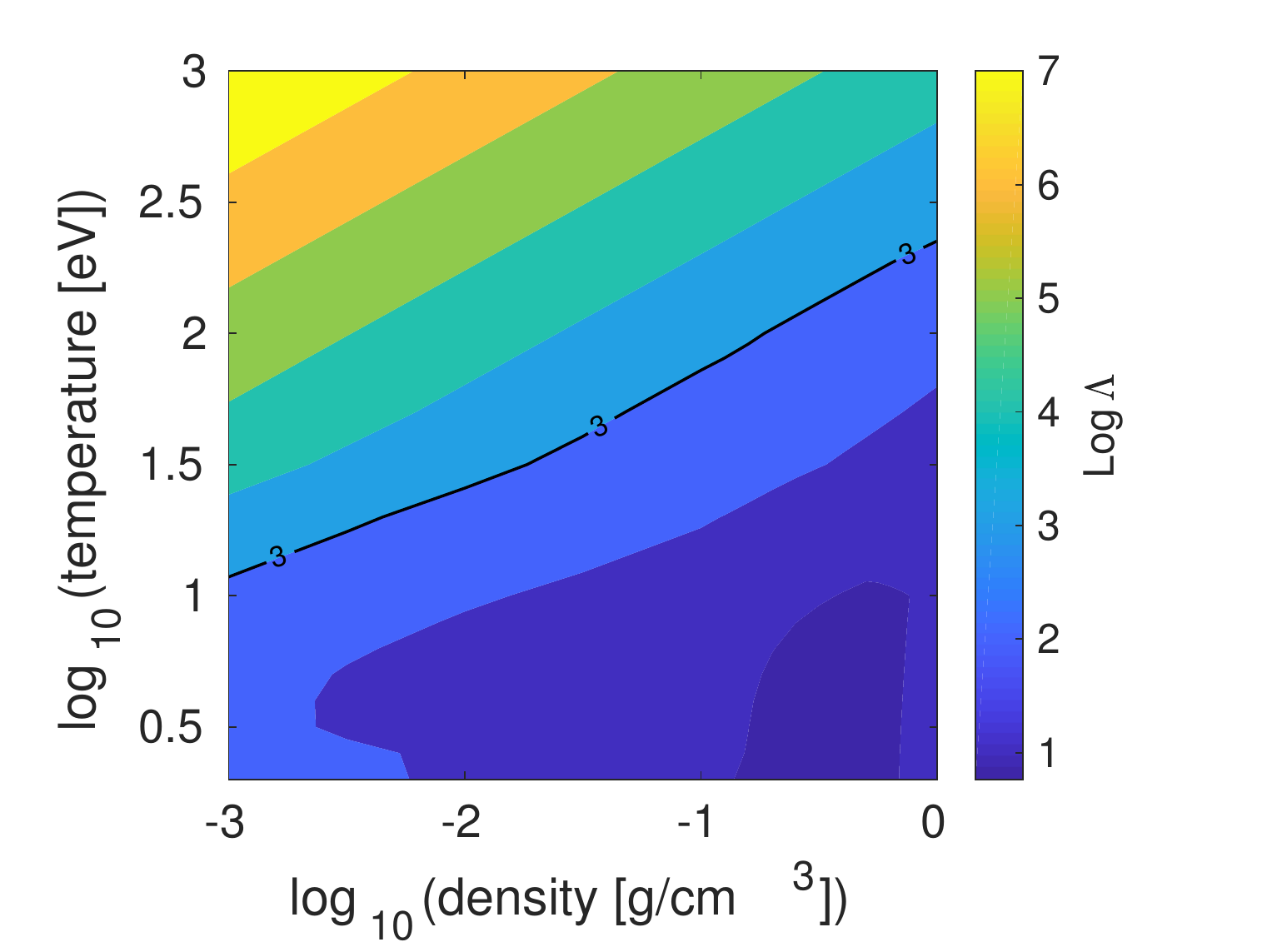}
  \end{subfigure}
  \begin{subfigure}[]{} 
    \includegraphics[scale=0.6]{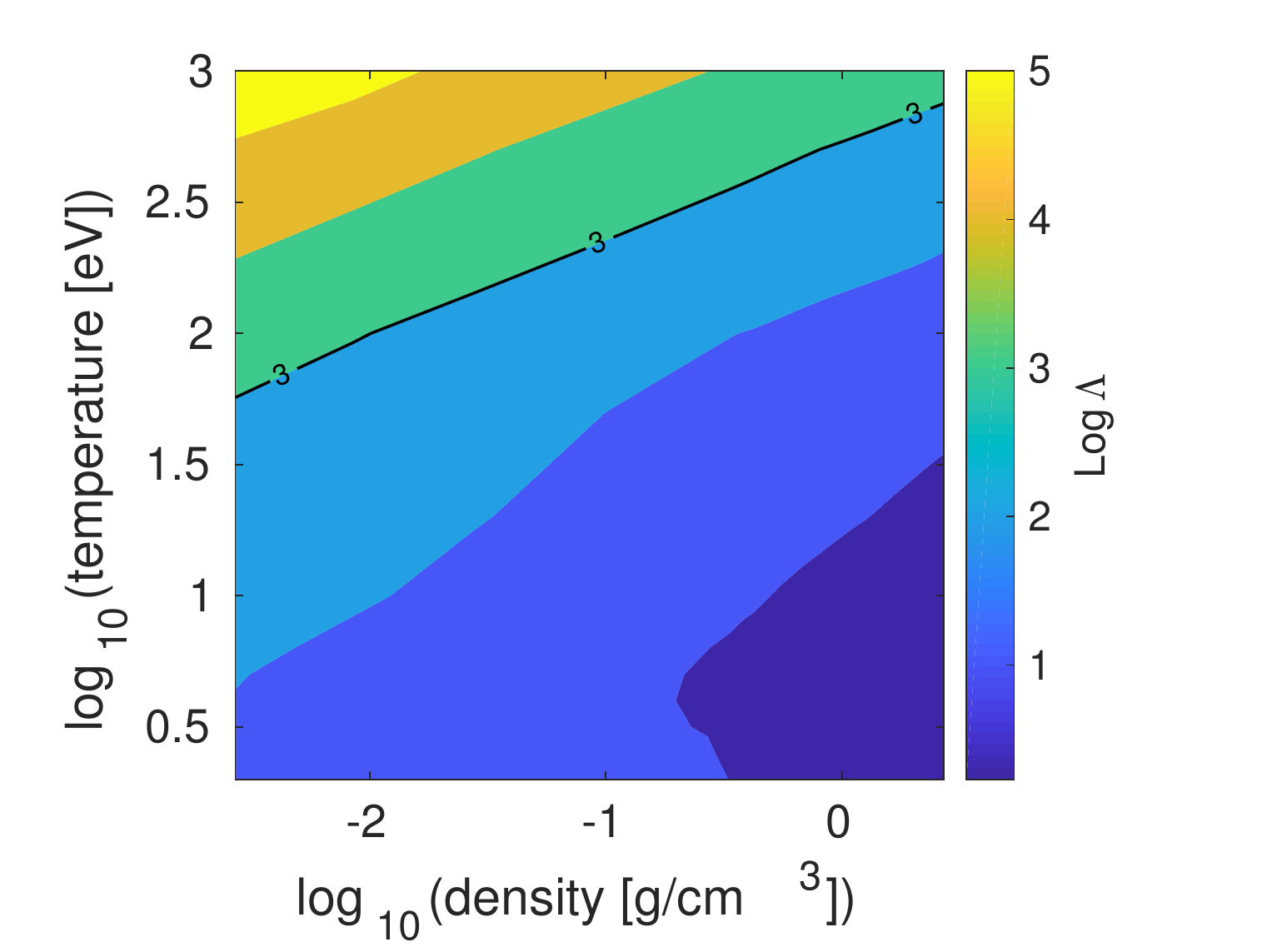}
  \end{subfigure}
  \begin{subfigure}[]{} 
    \includegraphics[scale=0.6]{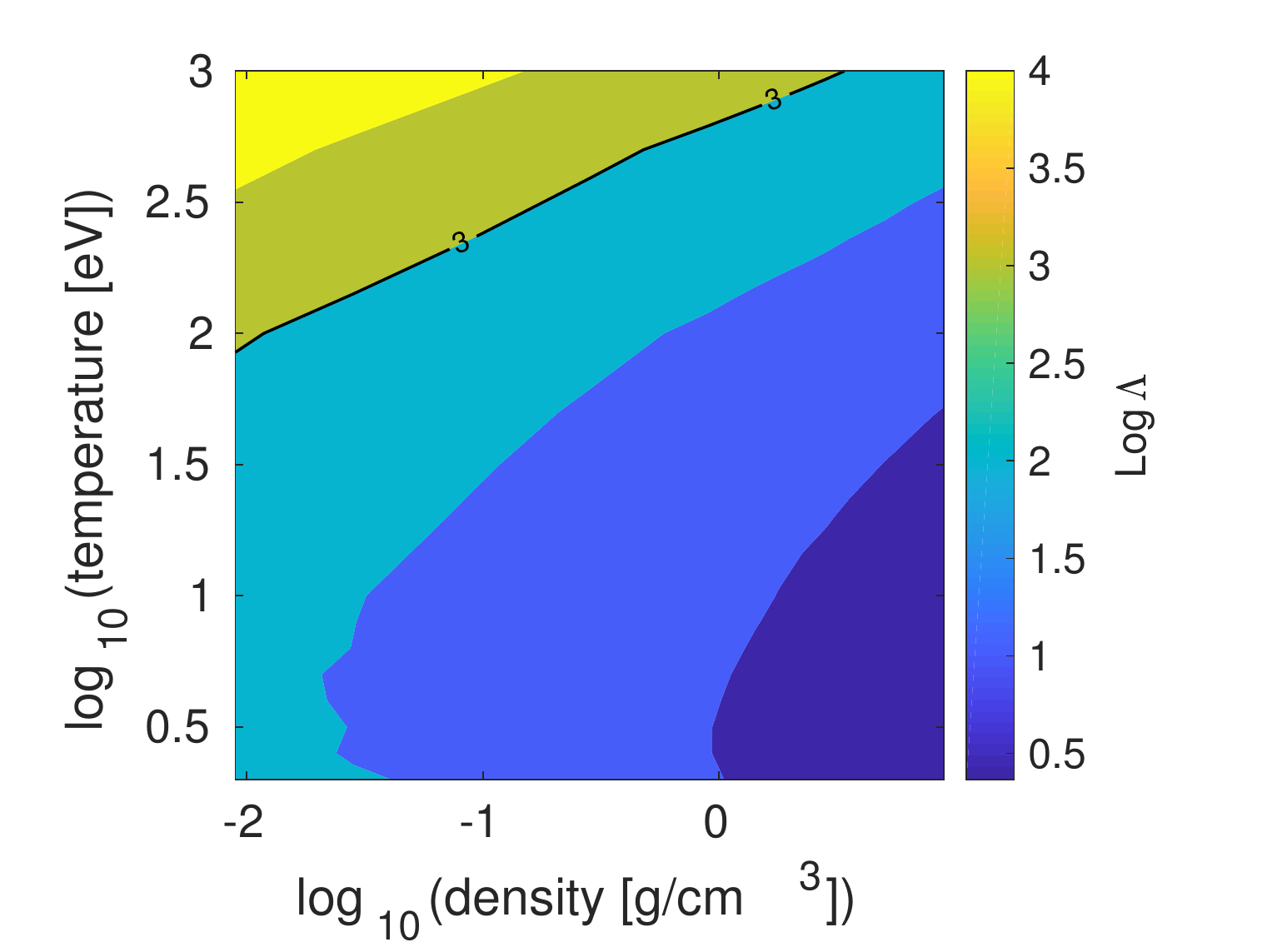}
  \end{subfigure}
  \caption{Contour plot of $\log\Lambda$ for (a) hydrogen, (b) aluminum and (c) copper.} 
  \label{fig_con}
\end{figure}

The value of the Coulomb Log is sometimes used as a metric for the
validity of the approximation itself.  For example, LM allow
a minimum value of $\log\Lambda$ = 2.0.  In figure \ref{fig_logl}
the value of $\log\Lambda$ is plotted against the ratio of
the quantum result to the classical result (without the Born correction).
The results correspond to all cases considered in figures \ref{fig_hbe} to
\ref{fig_aglu}, for the full range of densities and temperatures.
An interesting behavior is observed; for all materials considered a similar
trend in accuracy is seen.  This semi-universal behavior
suggests the inaccuracy of the classical method is caused by the same
physics, irrespective of the material, and perhaps that the classical
approximation could be further improved, though this is not attempted here.

For $\log\Lambda > 3$ the classical method is accurate to $\pm 10\%$.  This
confirms the use of $\log\Lambda$ as a indicator of accuracy.  For 
$\log\Lambda \approx 2$ the error is $\sim 50 \%$.  For smaller values
of $\log\Lambda$ there first appears a crossing point, then the classical
method grossly underestimates the quantum result.

The classical $\log\Lambda$ approach 
depends on knowledge of the average ionization $\bar{Z}$.  
For figures \ref{fig_hbe} to \ref{fig_logl} this was
determined using the same model as was used to generate
the potential $V^{\sss MF}$.  This is a quantum mechanical
DFT based model that takes a few minutes per density and
temperature point to run. A much faster and more widely
used model is the Thomas-Fermi cell (TFC) model \cite{feynman}.
This model is used extensively to give a quick estimate of
the plasma equation of state and ionization.  It is very
computationally stable and efficient, taking only a fraction
of a second to return a result.  The price of this speed
is physical accuracy.  

In figure \ref{fig_logl_tf} 
the full quantum results are compared with the classical
$\log\Lambda$ method using the TFC model for $\bar{Z}$.  Different
options are available for the ionization definition
using the TFC model \cite{murillo13}.  We have used
the electron density at the edge of the cell, which
is equal to the electron density in zero potential
with a chemical potential determined by the model.
Somewhat surprisingly, the comparison in figure
\ref{fig_logl_tf} shows that the TFC ionization
gives similar agreement to using $\bar{Z}$ from the full quantum calculation
as seen in figure \ref{fig_logl}.  This can be explained
by the relatively high temperatures considered here (10,
100 and 1000 eV), where the TFC model is more accurate.
This is a useful result.  It means that for $\log\Lambda$
$>$3 the TFC model can be used to rapidly estimate the
DC conductivity of dense plasmas to $\pm$ 10\% accuracy.

What region of phase space does $\log\Lambda > $ 3 correspond to?  Of course
this depends on the material.  In figure \ref{fig_con} contour plots of $\log\Lambda$
versus temperature and density for hydrogen, aluminum and copper are shown.
Clearly, as Z increases the region where $\log\Lambda >$ 3 shrinks, leaving
only the high temperature, low density region.  This region corresponds
to the least degenerate plasma.  Assuming an accuracy better than $\pm 10\%$ is required for
the DC conductivity, it is clear that the classical $\log\Lambda$
results cover only a limited region of phase space, and the more accurate
quantum approach must be used elsewhere. 

\begin{figure}
\begin{center}
\includegraphics[scale=0.30]{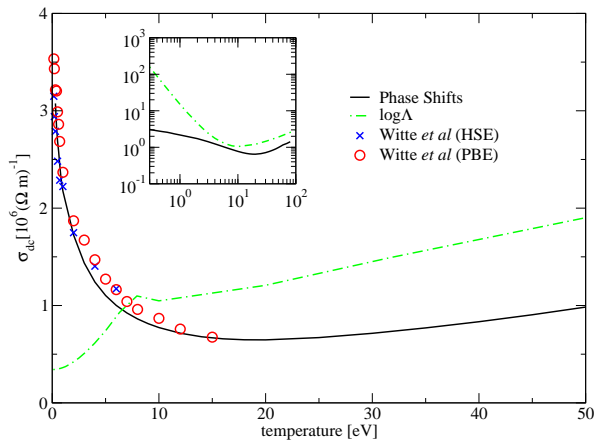}
\end{center}
\caption{(Color online) 
Electrical conductivity of solid density aluminum.  The full
quantum mechanical calculation (Phase Shifts) and the
classical method ($\log\Lambda$) are compared
to Kubo-Greenwood DFT-MD calculations of Witte {\it et al} \cite{witte18}.
The inset compares the fully quantum mechanical calculations to
$\log\Lambda$ results that do not use degeneracy corrections (see text).
}
\label{fig_witte}
\end{figure}

As a final example to highlight this result, in figure \ref{fig_witte}
these models are compared to recent Kubo-Greenwood, DFT-MD simulation
results \cite{witte18} for solid density aluminum in the relatively low temperature 
regime. There are two sets of DFT-MD results corresponding to two different
exchange and correlation potentials (PBE \cite{pbe} and HSE \cite{hse}).  The full
quantum model agrees well with both sets of DFT-MD results. The agreement is
not perfect but given the very different models it is very encouraging.  In contrast
the classical $\log\Lambda$ result is qualitatively and quantitatively different.
The change in behavior at $\sim 8$ eV is due to the degeneracy corrections to the
impact parameters.  In the inset the result without these degeneracy corrections
is shown, i.e. replacing equations (\ref{ldh}) and (\ref{vbar})
\begin{equation}
\lambda_{\sss DH}^{-2} = 
\frac{4\pi \bar{n}_e^0 e^2 }{kT}
+ 
\frac{4\pi \bar{Z}^2 n_i^0 e^2}{kT}
\end{equation}
and
\begin{equation}
\bar{v} = v_{rms}
\end{equation}
Clearly, the degeneracy corrections do not yield accurate conductivities and they provide
a only a marginal improvement over the uncorrected impact parameters.

\section{Conclusions}
A comparison of a quantum mechanical calculation for a realistic scattering potential
to the classical $\log\Lambda$ approach has revealed that the classical
method is accuracy to $\pm 10\%$ when $\log\Lambda >3$.  The classical 
method requires an ionization model and it was found that the
widely used and inexpensive Thomas-Fermi-Cell (TFC) model \cite{feynman}
provides a sufficiently accurate ionization estimate where the $\log\Lambda$
method is also accurate (i.e. $\log\Lambda > 3$).

The thermodynamic regime which corresponds to $\log\Lambda > 3$ depends strongly
on the material.  For increasing $Z$ this regime moves to higher temperatures and lower
densities.  This implies that, if accuracies in electrical conductivity better
than $\pm 10\%$ are required, then the $\log\Lambda$ method can only be used if
$\log\Lambda > 3$ and in general one must generate tables on data that can later be
interpolated.

It is noted that other transport coefficients such as thermal conductivity, 
electron-ion temperature relaxation
and stopping power also rely on the electron momentum transport cross section.
Hence, while this work is limited to electrical conductivity, the conclusions
are expected to translate to these other processes, though perhaps in a 
non-trivial way.

This study is limited in a number of ways; only electron-ion
collisions have been considered, the effect of electron-electron collisions on the conductivity
has not been studied here;  the plasma has been assumed to be made up of 
an ensemble of identical pseudo-atoms that all have the same average ionization; and only
pure plasmas have been considered, i.e. mixtures are not considered.
However, none of these limitations affect the conclusions since they are
common to both the classical and quantum approaches.
They do, however, point to future directions for improvement and testing.

\section*{Acknowledgments}
The author thanks B. Witte and S. X. Hu for providing their DFT-MD results.
This work was performed under the auspices of the United States Department of Energy under contract DE-AC52-06NA25396.

\bibliographystyle{unsrt}
\bibliography{phys_bib}

\end{document}